%% file: dirac5.tex
\documentclass[aps,nobibnotes,nofootinbib,twocolumn,nopacs,showkeys,superscriptaddress,groupedaddress]{revtex4}

\usepackage{color,stmaryrd,amsmath,amsthm,amsfonts,amssymb,times,bbm,graphicx,tikz,verbatim}

\include{kkmacros}

\begin{document}

\title[]{The Dirac equation as a quantum walk:\\ higher dimensions, observational convergence}

\author{Pablo Arrighi}
\affiliation{LIG, Universit\'e Joseph Fourier, Grenoble, France}
\affiliation{Universit\'e de Lyon, LIP, 46 all\'ee d'Italie, 69008 Lyon, France}  
\email{pablo.arrighi@imag.fr}

\author{Vincent Nesme}
\affiliation{LIG, Universit\'e Joseph Fourier, Grenoble, France}  
\email{vincent.nesme@imag.fr}

\author{Marcelo Forets}
\email{mforets@fing.edu.uy}
\affiliation{LIG, Universit\'e Joseph Fourier, Grenoble, France}

\keywords{Friedrichs symmetric hyperbolic systems, Quantum Walk, Quantum Lattice Gas Automata, Quantum Computation, Trotter-Kato, Baker-Campbell-Thomson, Operator splitting, Lax theorem}

\begin{abstract}
The Dirac equation can be modelled as a quantum walk, with the quantum walk being: discrete in time and space (i.e. a unitary evolution of the wave-function of a particle on a lattice); homogeneous (i.e. translation-invariant and time-independent), and causal (i.e. information propagates at a bounded speed, in a strict sense). This quantum walk model was proposed independently by Succi and Benzi, Bialynicki-Birula and Meyer: we rederive it in a simple way in all dimensions and for hyperbolic symmetric systems in general. We then prove that for any time $t$, the model converges to the continuous solution of the Dirac equation at time $t$, i.e. the probability of observing a discrepancy between the model and the solution is an $O(\varepsilon^2)$, with $\varepsilon$ the discretization step. At the practical level, this result is of interest for the quantum simulation of relativistic particles. At the theoretical level, it reinforces the status of this quantum walk model as a simple, discrete toy model of relativistic particles.
\end{abstract}

\maketitle

\section*{Introduction}

{\em The Dirac equation.} This PDE is the main equation for describing the behaviour of relativistic quantum particles. For a free fermion of mass $m$, it takes the form (in Planck units $\hbar=c=1$):
\begin{align}
\ii\partial_0 \ket{\psi} &= D\ket{\psi},\quad \textrm{with}\quad
D=m\alpha^0-\ii\sum_j \alpha^j\partial_j \label{eq:DiracFull}
\end{align}
where: 
\begin{itemize}
\item Latin index $j$ spans the spatial dimensions $1\ldots n$ 
whereas Greek indices $\mu,\nu$ will span the space-time dimensions $0\ldots n$. 
\item $\ket{\psi}$ is a space-time wave-function from $\mathbb{R}^{n+1}$ to $\mathbb{C}^d$, with $d$ a number that depends on $n$, whereas $\ket{\phi}$ will denote a space-like wave-function from $\mathbb{R}^{n}$ to $\mathbb{C}^d$.
\item  The $(\alpha^\mu)$ are $d\times d$ hermitian matrices which must verify $\{\alpha^\mu,\alpha^\nu\}=2\delta_{\mu\nu}\Id$, i.e. they square to the identity and pairwise anticommute. The notation $A\ket{\psi}$, with $A$ a $d\times d$ matrix, stands for the function that maps $(x_\mu)\in \mathbb{R}^{n+1}$ to $A\ket{\psi(\ldots x_\mu\ldots)}$.
\item  The notation $\ket{\psi(x_\mu)}$ stands for the function that maps $(x_\nu)_{\nu\neq\mu}\in \mathbb{R}^{n}$ to $\ket{\psi(\ldots x_\mu\ldots x_\nu \ldots )}$, e.g. we may write $\ket{\phi}=\ket{\psi(x_0=0)}$ for the initial state. The notation $\partial_\mu\ket{\psi}$ stands for the partial derivative with respect to the $\mu$-th coordinate.
\end{itemize}

\noindent {\em Discretization.} For the purpose of quantum simulation (on a quantum device) as envisioned by Feynman \cite{FeynmanQC}, or for the purpose of exploring the power and limits discrete models of physics, we may wish to discretize the Dirac equation. There are (at least) two obvious directions one could follow. First, through finite-difference methods one gets (where $\tr{\mu,\varepsilon}$ denotes translation by $\varepsilon$ along the $\mu$-axis):
\begin{align*}
\ket{\psi(x_0+\varepsilon)} &= (\Id-\ii \varepsilon D_\varepsilon) \ket{\psi(x_0)},\\
\textrm{with}\quad &D_\varepsilon =m\alpha^0-\ii\sum_j \alpha^j\frac{\tr{j,\varepsilon}-\Id}{\varepsilon}, \\
&(\tr{\mu,\varepsilon}\ket{\psi})(x_\mu) = \ket{\psi(x_\mu+\varepsilon)}
\end{align*}
The problem with this crude approach is that $(\Id-\ii \varepsilon D_\varepsilon)$ does not conserve the $||.||_2$-norm, in general.  From the point of view of numerical simulation, this means one has to check the model's convergence and stability. From the point of view of quantum simulation this simply bars the model as not implementable on a simulating quantum device. From the point of view of discrete toy models of physics, this means that the model lacks one of the fundamental, guiding symmetries: unitarity.

The second approach would be integrating exactly the original Dirac equation, and expressing $\ket{\psi(x_0+\varepsilon)}$ as a function of $\ket{\psi(x_0)}$.  The transformation would be unitary, but it is unclear how to discretize space.

\noindent  {\em The Dirac Quantum Walk.} In \cite{BenziSucci,Bialynicki-Birula,MeyerQLGI}, the Dirac equation is modelled as a Quantum Walk, i.e. a dynamics having the following features:
\begin{itemize}
\item The spacetime is a discrete grid;
\item The evolution is unitary; 
\item It is homogeneous, i.e. translation-invariant and time-independent;
\item It is causal, i.e. information propagates strictly at a bounded speed.
\end{itemize} 
In fact, \cite{MeyerQLGI} is considered to be one of the seminal papers about Quantum Walks \cite{Kempe}.\\
In numerical analysis, in order to evaluate the quality of a numerical scheme model, two main criteria are used. The first criterion is consistency, a.k.a. accuracy. Intuitively it demands that, after an $\varepsilon$ of time, the discrete model approximates the solution to a given order of $\varepsilon$. \\
Consistency of the $(1+1)$-dimensional Dirac Quantum Walk has been argued in \cite{MeyerQLGI}, and for the $(1+1)$-dimensional massless case in \cite{IndiansDirac}. It has been observed numerically in $(1+1)$-dimensions in \cite{LoveBoghosian} and in $(3+1)$-dimensions in \cite{PalpacelliThesis,LapitskiDellar,LapitskiDellarPalpacelliSucci}. It has been proved in $(1+1)$-dimensions in \cite{StrauchShrodinger,StrauchPhenomena,DAriano}. \\
{\em In this paper we provide a simple and formal derivation of the consistency of the Quantum Walk model of the Dirac equation, which works in full generality: we do not limit ourselves to the massless case, nor to the $(1+1)$-dimensional case.}

The second criterion is convergence. Intuitively it demands that, after an arbitrary time $x_0$, and if $\varepsilon$ was chosen small enough, the discrete model approximates the solution to a given order of $\varepsilon$. This criterion is stronger\footnote{Of course convergence implies consistency, but the converse does not always hold. Indeed, consistency means that making $\varepsilon$ small will increase the precision of the simulation of an $\varepsilon$ of time step. But it will also increase the number of time steps $k=x_0/\varepsilon$ which are required in order to simulate an $x_0$ of time evolution. Depending upon whether the two effects compensate, convergence may or may not be reached.}. Convergence has been observed numerically in $(3+1)$-dimensions in \cite{PalpacelliThesis,LapitskiDellar,LapitskiDellarPalpacelliSucci}. It has been proved in $(1+1)$-dimensions in \cite{StrauchShrodinger,StrauchPhenomena}. {\em In this paper, we provide a simple and formal derivation of convergence, which works in full generality: we do not limit ourselves to the massless case, nor to the $(1+1)$-dimensional case.}

The difficulty to analyse the $(3+1)$-dimensional Dirac Quantum Walk is mentioned in \cite{StrauchCTQW,StrauchPhenomena,BoghosianTaylor1,ChaMeyer}. Our approach is  based upon techniques such as: Sobolev spaces; Symmetric hyperbolic systems; Operator splitting, the Lax theorem. {\em We also address the question of the discretization of the input wavefunction $\ket{\phi}$. Altogether we prove that for any time $x_0$ and a sufficiently regular initial condition $\ket{\phi}$, the probability of observing a discrepancy between the iterated walk $\RC(W_\varepsilon^{x_0/\varepsilon}\D(\ket{\phi}))$ and the solution of the Dirac equation $\ket{\psi(x_0)}=T(x_0)\phi$, goes to zero, quadratically, as the discretization step $\varepsilon$ goes to zero.}

\noindent  {\em Other related works.}  The non-relativistic Dirac to Shr\"odinger limit of the Dirac Quantum Walk is studied in \cite{BenziSucci,StrauchCTQW,StrauchShrodinger,BoghosianTaylor2}. Decoherence, entanglement and Zitterbewegung are studied in \cite{LoveBoghosian,StrauchPhenomena}. Refinements aimed at numerical simulations and accounting for the Maxwell-Dirac equations or the time-dependent Dirac equation are given in \cite{LorinBandrauk,HJMSZ,FillionLorinBandrauk}. Algorithmic applications of the Dirac Quantum Walk are studied in \cite{ChildsGoldstone}. First principles derivations in $(1+1)$ and $(3+1)$-dimensions are provided in \cite{DAriano,DAriano3D}.

The ideas behind the $(1+1)$-dimensional Dirac Quantum Walk can be traced back to Feynman's relativistic checkerboard \cite{Bateson2012}, although early models where not unitary \cite{Kauffman} and sometimes continuous-time Ising-like \cite{Gersch_chessboard}. In $(2+1)$-dimensions, continuous-time models over the honeycomb lattice have been conceived in order to model electron transport in graphene \cite{japanese_honeycomb}.

In \cite{MolfettaDebbasch} the authors define a discrete-time quantum walk modelling the $(3+1)$-dimensional Dirac equation. It is not homogeneous: neither is it translation-invariant, nor time-independent. But it reproduces samplings of the continuous solution exactly.

We start with informal derivations in $(2+1)$ and $(3+1)$-dimensions (Section \ref{sec:informal}). We recall well-posedness results for the Dirac equation (Section \ref{sec:wellposedness}), and continue with the formal analysis of the model, proving: consistency, stability and convergence (Sections \ref{sec:consistency}, \ref{sec:stability} and \ref{sec:convergence}). Finally, we discuss space discretization and other considerations such as generalizations and observational equivalence (Sections\ref{sec:spacediscretization} and \ref{sec:considerations}).

\section{Informal derivations}\label{sec:informal}

\noindent A standard representation of the $(2+1)$-dimensional Dirac equation is:
\begin{align}
\ii\partial_0 \ket{\psi} &= D\ket{\psi}\quad \textrm{with}\quad D=m\sigma^2 -\ii\sigma^1 \partial_1 -\ii\sigma^3 \partial_2\label{eq:Dirac2D}
\end{align}
and $(\sigma^\mu)$ the Pauli matrices (with $\sigma^0$ the identity).
Now, intuitively,
\begin{align}
\tr{\mu,\varepsilon}\ket{\psi}=(\Id+\varepsilon\partial_\mu)\ket{\psi}+O(\varepsilon^2).
\label{eq:approxshift}
\end{align}
but this statement and its hypotheses will only be made formal and quantified in later sections. Meanwhile, substituting Eq. (\ref{eq:Dirac2D}) into Eq. (\ref{eq:approxshift}) for $\mu=0$ yields:
\begin{align*}
\tr{0,\varepsilon}&= (\Id-\ii\varepsilon D)+O(\varepsilon^2)\\
&= (\Id-\ii\varepsilon m \sigma^2)(\Id -\varepsilon \sigma^1 \partial_1)(\Id -\varepsilon \sigma^3 \partial_2)+O(\varepsilon^2)\\
&= \exp\pa{-\ii\varepsilon m \sigma^2}H(\Id -\varepsilon \sigma^3 \partial_1)H(\Id -\varepsilon \sigma^3 \partial_2)+O(\varepsilon^2)
\end{align*}
since $\sigma^1 = H \sigma^3 H$ with $H$ the Hadamard gate.\\
Using the definition of $\sigma^3$, Eq. (\ref{eq:approxshift}), and taking the convention that ${\mathbb{C}}^2$ is spanned by the orthonormal basis $\{|l\rangle / l \in \{-1,1\} \}$, we get:
\begin{align*}
\tr{0,\varepsilon}&= C_\varepsilon HT_{1,\varepsilon}HT_{2,\varepsilon}+O(\varepsilon^2)\\
\textrm{with}\quad C_\varepsilon  &=\exp \pa{-\ii\varepsilon m \sigma^2}\\
\textrm{and}\quad T_{j,\varepsilon}&=\sum_{l\in\{-1,1\}} |l\rangle\bra{l}\tr{j,l\varepsilon}.
\end{align*}
Overall, we have:
\begin{align*}
\ket{\psi(x_0+\varepsilon)} &= W_\varepsilon\ket{\psi(x_0)}+O(\varepsilon^2)\\
\textrm{with}\quad W_\varepsilon&=C^{\varepsilon}HT_{1,\varepsilon}HT_{2,\varepsilon}
\end{align*}
where the $T$ matrices are partial shifts. This Dirac Quantum Walk \cite{BenziSucci,Bialynicki-Birula,MeyerQLGI} models the $(2+1)$-dimensional Dirac equation. It has a product form. Such `alternate quantum walks' have the advantage of using a two-dimensional coin-space instead of a four-dimensional coin-space: fewer resources are needed for their implementation \cite{McGettrick}. It is still just one quantum walk, i.e. a translation-invariant causal unitary operator.

From $(2+1)$ to $(3+1)$-dimensions the Dirac equation changes form, the spin degree of freedom goes to degree four. The equation is: 
\begin{align*}
\ii\partial_0 \ket{\psi} &= D\ket{\psi}\quad \textrm{with}\\
D&= m(\sigma^2\otimes\sigma^0) +\ii \sum_j(\sigma^3\otimes\sigma^j)\partial_j
\end{align*}
Indeed, one can check that the matrices $\sigma^2\otimes\sigma^0$ and $(-\sigma^3\otimes\sigma^i)$ are hermitian, that they square to the identity, and that they anticommute.
Using the definition of $\sigma^3$, Eq. (\ref{eq:approxshift}), and taking the convention that ${\mathbb{C}}^4$ is spanned by the orthonormal basis $\{ |r, l\rangle\,/\, r,l \in \{-1,1\} \}$:
\begin{align*}
\left(  \Id+\varepsilon (\sigma^3\otimes\sigma^3)\partial_3 \right) \ket{\psi}&=T_{3,\varepsilon}\ket{\psi}+O(\varepsilon^2)\\
\textrm{with}\quad T_{j,\varepsilon}=&\sum_{r,l\in\{-1,1\}} |r,l\rangle\bra{r,l}\tr{j,rl\varepsilon}.\end{align*}
Similarly,
\begin{align*}
\left( \Id+\varepsilon (\sigma^3\otimes\sigma^2)\partial_2 \right) \ket{\psi} &= (\Id\otimes F) T_{2,\varepsilon} (\Id\otimes F^\dagger) \ket{\psi}+O(\varepsilon^2)\\
\textrm{as}\quad \sigma^2 = &F \sigma^3 F^\dagger\\
\textrm{with}\quad F = &R_{\frac{\pi}{2}}H = 
\left(\begin{array}{cc}
1/\sqrt{2} &1/\sqrt{2}\\
\ii/\sqrt{2} &-\ii/\sqrt{2}
\end{array}\right).
\end{align*}
Likewise,
\begin{align*}
\left( \Id+\varepsilon (\sigma^3\otimes\sigma^1)\partial_1 \right) \ket{\psi}  &= (\Id\otimes H) T_{1,\varepsilon} (\Id\otimes H) \ket{\psi}+O(\varepsilon^2)\\
\textrm{as}\quad \sigma^1 = &H\sigma^3 H.
\end{align*}
Finally, let $C_\varepsilon=\exp\pa{-\ii\varepsilon m (\sigma^2\otimes\sigma^0)}$. 
We have:
\begin{align*}
\ket{\psi(x_0+\varepsilon)}&=W_\varepsilon\ket{\psi(x_0)}+O(\varepsilon^2)\\
\textrm{with}\quad W_\varepsilon&=C_{\varepsilon}(\Id\otimes H)T_{1,\varepsilon}(\Id\otimes HF)T_{2,\varepsilon}(\Id\otimes F^\dagger) T_{3,\varepsilon}
\end{align*}
where the $T$ matrices are partial shifts. This is the $(3+1)$-dimensional Dirac Quantum Walk. We now move on to the formal analysis of the model.

\section{Well-posedness}\label{sec:wellposedness}

Numerical analysis is mostly about finding discrete models to approximate the continuous solutions of a well-posed Cauchy problem.\\
Here, the Cauchy problem is to find the solution $\ket{\psi}$ given $\ket{\psi(0)}$ and $\ii\partial_0 \ket{\psi} = D\ket{\psi}$. Cauchy problems are well-posed if and only if the solution exists, is unique, and depends continuously upon $\ket{\psi(0)}$. Since the Dirac equation is a symmetric hyperbolic system, the problem is known \cite{Fattorini} to be well-posed for the Sobolev space $\HS{s}$, with $s\geq 0$ of the functions for which the $||.||_\HN{s}$-norm is finite. This Sobolev norm
\begin{align*}
||\ket{\phi}||_\HN{s}&=\sqrt{ \int_{\mathbb{R}^n} (1+m^2+|| k||^2)^{s}||\ket{\hat{\phi}( k)}||^2 \dd k},
\end{align*}
and the well-posedness result are discussed in Appendix \ref{sec:Sobolev}. Notice that $\HS{0}$ is the usual $\LS$. Notice also that the Sobolev norm involves an integral in Fourier space. For this reason, and because the Dirac operator is just a pointwise multiplication in Fourier space, most of our derivations will use it. Conventions and basic facts about Fourier space are given in Appendix \ref{sec:Fourier}.

\section{Consistency}\label{sec:consistency}

In numerical analysis, in order to evaluate the quality of a numerical scheme model, the first criterion is {\em consistency}, a.k.a. accuracy. Intuitively it demands that, after an $\varepsilon$ of time, the discrete model approximates the solution to a given order of $\varepsilon$.\\
Formally, say a Cauchy problem is well-posed on $X$, with $Y$ a dense subspace of $X$. The discrete model $W_\varepsilon$ is consistent of order $r$ on $Y$ if and only if there exists $C$ such that for any solution $\ket{\psi}$ with $\ket{\psi(x_0=0)}\in Y$, for all $\varepsilon\in\mathbb{R}^+$, we have
$$||W_\varepsilon\ket{\psi(0)}-\ket{\psi(\varepsilon)}||_{X}=\varepsilon^{r+1}C||\ket{\psi(0)}||_Y.$$ This is what we will now prove: that for $s\geq 0$, $r=1$, $X=\HS{s}$ and $Y=\HS{s+2}$, there exists $C$ such that for all $\ket{\phi}$, $\varepsilon$:
$$ ||W_{\varepsilon}\ket{\phi} -  T(\varepsilon)\ket{\phi}||_{\HN{s}}\leq \varepsilon^2 C ||\ket{\phi}||_\HN{s+2},$$
with $\ket{\phi}=\ket{\psi(0)}$, $T(\varepsilon)\ket{\phi}=\ket{\psi(\varepsilon)}$, i.e. $T(\varepsilon)=\tau_{0,\epsilon}$ is the continuous solution's time evolution operator. 

We work on Fourier space and see $\hat{W}_\varepsilon( k)$ with fixed $ k$ as a function of the real-value $\varepsilon$.
First, observe that the quantum walk operator can generally be written as (we sometimes omit the $ k$ dependence in the notations of this section):
\begin{equation}
\hat{W}_\varepsilon = \prod_{\mu} e^{-\ii \varepsilon \hat{A}_\mu }. \label{eq:QWOperator}
\end{equation}
With $\hat{A}_\mu$ hermitian, $|||\hat{A}_0|||_2=m$, $|||\hat{A}_j|||_2 =  k_j$, $\hat{A}_\mu$ hermitian and $\sum_\mu \hat{A}_\mu = \hat{D} $ (see Appendix \ref{sec:Fourier} for further details). For instance, in $(2+1)$-dimensions, $\hat{A}_0$ is equal to $m\sigma^2$, $\hat{A}_1$ is equal to $ k_1\sigma^1$ and $\hat{A}_2$ is equal to $ k_2\sigma^3$ (see Appendix \ref{sec:Fourier} for further details). 

As $\hat{W}_\varepsilon( k)$ is a matrix whose elements are products of trigonometric functions and exponentials, its entries are ${\cal{C}}^\infty$ functions (on the variable $\varepsilon$). We will denote $\partial_\varepsilon$ the derivative with respect to variable $\varepsilon$ in each entry. Observe that $ \hat{W}_{0} = \Id$.

Now we will calculate the first and second order derivatives making use of Eq. (\ref{eq:QWOperator}). 
For the first order derivative we have
\begin{small}
\begin{align*}
\left(\partial_\varepsilon \hat{W}_\varepsilon\right)_\varepsilon &=  \sum_{\mu} \left( \prod_{\kappa < \mu} e^{-\ii \varepsilon \hat{A}_{\kappa}} \right)  \left(-\ii \hat{A}_\mu \right) \left( \prod_{\kappa \geq \mu}   e^{-\ii \varepsilon \hat{A}_{\kappa}} \right)
\end{align*}
\end{small}

Evaluating at $\varepsilon=0$, 
\begin{align*}
\left(\partial_\varepsilon \hat{W}_\varepsilon\right)_{\varepsilon=0} = -\ii \hat{D} 
\end{align*}

For the second order derivative, we have:
\begin{small}
\begin{align*}
&\left( \partial^2_\varepsilon \hat{W}_\varepsilon \right)_\varepsilon = -\sum_{\mu} \left( \prod_{\kappa < \mu} e^{-\ii \varepsilon \hat{A}_\kappa} \right)   \hat{A}_\mu^2 \left( \prod_{\kappa \geq \mu}   e^{-\ii \varepsilon \hat{A}_{\kappa}} \right) \\ &  - 2 \sum_{\nu < \mu} \left( \prod_{\kappa < \nu} e^{-\ii \varepsilon \hat{A}_\kappa}\right) \hat{A}_\nu \left( \prod_{\nu \leq \kappa < \mu}  e^{-\ii \varepsilon \hat{A}_{\kappa}} \right) \hat{A}_\mu \left( \prod_{\kappa \geq \mu}  e^{-\ii \varepsilon \hat{A}_{\kappa}} \right)
\end{align*}

\end{small}

\begin{align*}
||| \partial^2_\varepsilon \hat{W}_\varepsilon  |||_2 &\leq \sum_{\mu} |||\hat{A}_\mu^2 |||_2  + 2 \sum_{\nu < \mu} ||| \hat{A}_\nu |||_2 ~ |||\hat{A}_\mu|||_2 \\ & \leq \left(  \sum_{\mu} |||\hat{A}_\mu |||_2 \right)^2 
\\ &\leq (n+1) \sum_\mu |||\hat{A}_\mu |||_2^2 \\ &\leq (n+1) \gamma^2
\end{align*}
where we get to the preceding line using that for real numbers, $(x_0 + \cdots +x_n)^2 \leq (n+1)(x_0^2 +\cdots + x_n^2)$ and to the last line using $\gamma^2 = m^2 + || k||^2_2$. By application of Taylor's formula with the integral form for the remainder \cite{TaylorIF} to each entry of the matrix $\hat{W}_\varepsilon$, we get 
\begin{align*}
\hat{W}_\varepsilon &=  \Id  + \varepsilon \left(\partial_\varepsilon \hat{W}_\varepsilon\right)_{\varepsilon =0}  + \int_{0}^{\varepsilon} (\varepsilon -\eta) \left( \partial^2_\varepsilon \hat{W}_{\varepsilon } \right)_{\varepsilon = \eta} ~d\eta 
\end{align*}
and
\begin{align*}
\hat{T}(\varepsilon) &= e^{-\ii \varepsilon \hat{D}} = \Id -\ii \varepsilon \hat{D} \\ &+ \int_{0}^\varepsilon (\varepsilon -\eta) \left( -\hat{D}^2 e^{-\ii  \eta \hat{D}} \right) d\eta
\end{align*}

Let us define 
$$
\hat{R}_{\varepsilon} = \hat{W}_\varepsilon - \hat{T}(\varepsilon),
$$
whose operator norm can be bounded after substitution of the previous expressions and application of the triangular inequality, thus obtaining
\begin{align*}
|||\hat{R}_{\varepsilon}|||_2 &\leq \int_{0}^{\varepsilon} |\varepsilon -\eta |~||| \partial^2_\varepsilon \hat{W}_{\varepsilon } |||_2 ~d\eta \\ &+ \int_{0}^{\varepsilon}  |\varepsilon -\eta | ~ |||\hat{D}^2 e^{-\ii \varepsilon \eta \hat{D}}|||_2 d\eta \\ 
&\leq  \int_{0}^{\varepsilon}  (\varepsilon - \eta) (n+1) \gamma^2 d\eta + \int_{0}^{\varepsilon}  (\varepsilon - \eta) \gamma^2 d\eta \\ 
 &\leq \varepsilon^2 \gamma^2\left( 1+\frac{n}{2}\right)
\end{align*}
where we used that the eigenvalues of $\hat{D}$ are $\pm\gamma$ with $\gamma^2 = m^2 + || k||^2_2$, see Appendix \ref{sec:Fourier}. Substituting this result into the Sobolev norm, i.e.
\begin{align*}
& ||W_\varepsilon\ket{\phi} -  T(\varepsilon) \ket{\phi}||_{\HN{s}} =  \\ & \sqrt{ \int_{\mathbb{R}^n} (1+m^2 + || k ||^2)^{s} ||\hat{R}_\varepsilon \hat{\phi} ( k)||^2  \dd k } = \\
 & \sqrt{ \int_{\mathbb{R}^n} (1+m^2 + || k||^2)^{s}||\hat{R}_{\varepsilon}( k)\ket{\hat{\phi}}( k)||^2 \dd k }\\
&\leq \varepsilon^2 ~ C ~~ \sqrt{ \int_{\mathbb{R}^n} (1+m^2 + || k||^2)^{s+2} ||\hat{\phi}( k)||^2  \dd k } \\
&\leq \varepsilon^2 ~ C  ||\ket{\phi} ||_{\HN{s+2}}
\end{align*}

which is what we wanted to prove, $C$ being $1+\frac{n}{2}$.

\section{Stability}\label{sec:stability}

In numerical analysis, in order to evaluate the quality of a numerical scheme model, an intermediate criterion is {\em stability}. It demands the discrete model be a bounded linear operator. Thus, let us prove that for all $\ket{\phi}$, for all $s \geq 0$, we have $||W_\varepsilon \ket{\phi}||_\HN{s}=||\ket{\phi}||_\HN{s}$.
We proceed by applying the definition of Sobolev norm, which yields
\begin{align*}
||W_\varepsilon \ket{\phi}||^2_\HN{s}   
&= \int_{\mathbb{R}^n} (1+m^2+|| k||^2)^{s}||\FT (W_\varepsilon \ket{\phi}) ( k)||^2 \dd k \\
&= \int_{\mathbb{R}^n} (1+m^2+|| k||^2)^{s}|| (\hat{W}_\varepsilon \hat{\phi})( k)||^2 \dd k \\
&= \int_{\mathbb{R}^n} (1+m^2+|| k||^2)^{s}|| \hat{W}_\varepsilon ( k) \hat{\phi}( k)||^2 \dd k 
\end{align*}
where in the second to third lines we used the fact that as $W_\varepsilon$ is a translation-invariant unitary operator it is represented in Fourier space as a left multiplication by a unitary matrix $\hat{W}_\varepsilon( k)$, which depends on $ k$. See Appendix \ref{sec:Fourier} for this particular case, and for instance \cite{DaviesLinear} for the general case. We then have
\begin{align*}
||W_\varepsilon \ket{\phi}||^2_\HN{s} 
&= \int_{\mathbb{R}^n} (1+m^2+|| k||^2)^{s}|| \ket{\hat{\phi} ( k)}||^2 \dd k \\
&= ||\ket{\phi}||^2_\HN{s}
\end{align*}
Thus if $|||.|||_\HN{s}$ denotes the operator norm with respect to the norm $\HN{s}$, we have $|||W_\varepsilon|||_\HN{s}$ equal to one as requested.

\section{Convergence}\label{sec:convergence}

In numerical analysis, in order to evaluate the quality of a numerical scheme model, the most important criterion for quality is {\em convergence}. Intuitively it demands that, after an arbitrary time $x_0$, and if $\varepsilon$ was chosen small enough, the discrete model approximates the solution to a given order of $\varepsilon$. Fortunately, the Lax theorem \cite{LaxRichtmyer,ISEM-Lax} states that stability and consistency implies convergence. Unfortunately, as regards the quantified version of this result, the literature available comes in many variants, with various degrees of formalization, each requesting different sets of hypotheses. Thus, for clarity, we inline the proof here.

Formally, say a Cauchy problem is well-posed on $X$ and $Y$, with $Y$ a dense subspace of $X$. The discrete model $W_\varepsilon$ is convergent of order $r$ on $Y$ if and only if there exists $C$ such that for any solution $\ket{\psi}$ with $\ket{\psi(x_0=0)}\in Y$, for all $x_0\in \mathbb{R}^+$, $k\in\mathbb{N}$, we have:
$$||W_{\varepsilon_l}^l\ket{\psi(0)} - \ket{\psi(x_0)}||_X = \varepsilon_l^r x_0 C||\ket{\psi(0)}||_Y$$
with $\varepsilon_l=x_0/l$. 
This is exactly what we will now prove: that for $s\geq 0$, $r=1$, $X=\HS{s}$ and $Y=\HS{s+2}$, there exists $C$ such that for all $\ket{\phi}$, $\varepsilon$:
\begin{align*}
||W_{\varepsilon_l}^{l}\ket{\phi}-T(\varepsilon_l l)\ket{\phi}||_\HN{s}
&\leq \varepsilon_l  x_0 C ||\ket{\phi}||_\HN{s+2}
\end{align*}

Take $x_0\in \mathbb{R}^+$. Consider the sequence $(\varepsilon_l)$ such that $\varepsilon_l = x_0/l$. Because $T(\varepsilon_l l)=T(\varepsilon_l)^{l}$, and because 
\begin{align*}
\sum_{j=0}^{l-1} W_{\varepsilon_l}^{l-j}T(\varepsilon_l)^j - W_{\varepsilon_l}^{l-j}T(\varepsilon_l)^j&=0\\
\sum_{j=0}^{l-1} W_{\varepsilon_l}^{l-j}T(\varepsilon_l)^j - W_{\varepsilon_l}^{l-j-1}T(\varepsilon_l)^{j+1}&=W_{\varepsilon_l}^{l} - T(\varepsilon_l)^{l}
\end{align*}
We have:
\begin{align*}
W_{\varepsilon_l}^{l}\ket{\phi}-T(\varepsilon_l l)\ket{\phi} = 
\sum_{j=0}^{l-1} W_{\varepsilon_l}^{l-1-j}(W_{\varepsilon_l}-T(\varepsilon_l))T(\varepsilon_l)^j \ket{\phi}
\end{align*}
From consistency there exists $C$ such that for all $\ket{\phi}$,
$$ ||W_{\varepsilon_l} T(j\varepsilon_l)\ket{\phi} -  T(\varepsilon_l) T(j\varepsilon_l)\ket{\phi}||_{\HN{s}}\leq \varepsilon_l^2 C ||\ket{\phi}||_\HN{s+2}.$$
Hence,  
\begin{align*}
||W_{\varepsilon_l}^{l}\ket{\phi}-T(\varepsilon_l l)\ket{\phi}||_\HN{s}
&\leq \sum_{j=0}^{l-1} |||W_{\varepsilon_l}^{l-1-j}|||_{\HN{s}} \varepsilon_l^2 C ||\ket{\phi}||_\HN{s+2}\\
&\leq k \varepsilon_l^2 C ||\ket{\phi}||_\HN{s+2}
\leq \varepsilon_l  x_0 C ||\ket{\phi}||_\HN{s+2}
\end{align*}
as requested.

\section{Space Discretization}\label{sec:spacediscretization}

This paper aims at giving a quantum walk model $W_\varepsilon:\ell_2(\varepsilon \Z^n)^d\to \ell_2(\varepsilon \Z^n)^d$ of the Dirac equation. So far we explained how we can discretize time the Dirac equation, but in order to get a quantum walk, we need to discretize space as well. In a sense, this is already done since the walk operators $W_\varepsilon$ that we defined, although they take as input functions in $H^{s}(\R^n)^d$, can equally well be defined on $\ell_2(\varepsilon \Z^n)^d$, for the only shift operators involved in their definitions are multiples of the $T_{j,\varepsilon}$-s. The question remains, however, of what initial state we can feed our quantum walks, and how we are to interpret their output. Answering this question is the aim of this Section. One of the difficulties, in particular, is to construct, given $\phi\in\LS$, a $\D(\phi)\in\ell_2(\varepsilon \Z^n)^d$. That the discretized version of $\phi$ be normalized is essential so that the quantum simulation can be implemented on a quantum simulator, just like the unitarity of $W_\varepsilon$ was essential. This Section relies heavily on notations introduced in Appendix~\ref{sec:Fourier}.

\noindent {\em Discretization procedure.} We discretize by 
\begin{align*}
\D(\phi)&= \RN(\FS(\FT(\phi)|_{[-\frac{\pi}{\varepsilon},\frac{\pi}{\varepsilon}]^n}))
\end{align*}
Notice that 
$$\phi_{\textrm{LP}}=\FT^{-1}(\chi_{[-\frac{\pi}{\varepsilon},\frac{\pi}{\varepsilon}]^n}\FT(\phi)),$$
where $\chi_A$ denotes the indicator function of $A$, applies an ideal low-pass filter, and that 
$$\FS(\FT(\phi_{\textrm{LP}})|_{[-\frac{\pi}{\varepsilon},\frac{\pi}{\varepsilon}]^n})= \varepsilon^{n/2}\phi_{\textrm{LP}}|_{\varepsilon \Z^n}$$
is, up to a constant, the sampling of $\phi_{\textrm{LP}}$, see Appendix \ref{sec:Fourier}. $\D(\phi)$ is hence proportional to the function obtained by sampling $\phi$ after it has been low-pass filtered. Since $\FS$ and $\FT$ are unitary, the renormalization is by a factor of $\|\phi_{\textrm{LP}}\|_2^{-1}$. For it to be well-defined, we must check that $\phi_{\textrm{LP}}$ does have a non-zero norm. 

\noindent {\em Low-pass filtering.} For every $s\geq 0$, we have
\begin{small}
\begin{align*}
&\| \phi-\phi_{\textrm{LP}} \|_{\HN{s}} \\ &= \sqrt{\int_{\R^n\setminus [-\frac{\pi}{\varepsilon},\frac{\pi}{\varepsilon}]^n}  (1+m^2+\| k\|^{2})^{s} \|\hat\phi( k)\|^2 \dd k} \\
& = \sqrt{\begin{array}{l} \int_{\R^n\setminus [-\frac{\pi}{\varepsilon},\frac{\pi}{\varepsilon}]^n}  (1+m^2+\| k\|^{2})^{-2} \\ \hspace{1cm} (1+m^2+\| k\|^{2})^{s+2} \|\hat\phi( k)\|^2 \dd k \end{array}} \\
& \leq \varepsilon^{2}C' \| \phi \|_{\HN{s+2}} \quad \textrm{with}\quad  C'=\pi^{-2}. 
\end{align*}
\end{small}
This tells us two things. First, if  $\varepsilon^2 < \frac{\| \phi \|_2}{ C'\| \phi \|_{\HN{2}}}$, then $\phi_{\textrm{LP}} \neq 0$,  so it can be renormalized. 
Second, the loss induced by low-pass filtering is small, as needed below in order to bound the overall error.

\noindent {\em Reconstruction procedure.} We reconstruct by 
\begin{align*}
\RC(\tilde{\phi})&= \FT^{-1}(\FS^{-1}(\RN^{-1}(\tilde{\phi})))
\end{align*}
with the convention that $\FS^{-1}(\tilde{\phi})\in L^2([-\frac{\pi}{\varepsilon},\frac{\pi}{\varepsilon}]^n)^d$ is extended to $\LS$ by the null function on $\R^n\setminus[-\frac{\pi}{\varepsilon},\frac{\pi}{\varepsilon}]^n$, and the inverse renormalization is by a factor of $\|\phi_{\textrm{LP}}\|_2$. Notice that $$\phi_{\textrm{LP}}= \RC(\D(\phi_{\textrm{LP}}))$$
and that this reconstruction is equivalent to the Whittaker-Kotelnikov-Shannon formula (cf. \cite{pm62}, \cite{sampling_chinois} for the multidimensional case).

\noindent {\em Overall scheme.} 
Given a wave function $\phi$, we approximate $T(\varepsilon l)\phi$, the continuous evolution of $\phi$, by $\RC( W_{\varepsilon}^l (\D (\phi)))$ the reconstruction of the walk iterated on the discretization of $\phi$.  Let us bound the overall error. For all $\phi$ we have (renormalizations cancel out by linearity of $W_\varepsilon^l$):
\begin{align*}
&\RC(W_\varepsilon^l (\D(\phi)))\\
&= \FT^{-1}(\FS^{-1}(W_\varepsilon^l (\FS(\FT(\phi)|_{[-\frac{\pi}{\varepsilon},\frac{\pi}{\varepsilon}]^n}))))\\
&= \FT^{-1}(\FS^{-1}(W_\varepsilon^l (\FS(\FT(\phi_{\textrm{LP}})))))\\
&= \FT^{-1}(\FS^{-1}(W_\varepsilon^l (\varepsilon^{n/2}\phi_{\textrm{LP}}|_{\varepsilon \Z^n})))\\
&= \FT^{-1}(\FS^{-1}(\varepsilon^{n/2} W_\varepsilon^l (\phi_{\textrm{LP}})|_{\varepsilon \Z^n}))\\
&= \FT^{-1}(\FT(W_\varepsilon^l (\phi_{\textrm{LP}}))|_{[-\frac{\pi}{\varepsilon},\frac{\pi}{\varepsilon}]^n})\\
&= W_\varepsilon^l (\phi_{\textrm{LP}})
\end{align*}
where the preceding step comes from the last line of Appendix \ref{sec:Fourier}. Now, since $W_\varepsilon^l$ is unitary, we have 
\begin{align*}
&\|W_\varepsilon^l (\phi_{\textrm{LP}})-W_\varepsilon^l (\phi) \|_{\HN{s}}= \|\phi_{\textrm{LP}}-\phi \|_{\HN{s}}\leq \varepsilon^{2} C' \| \phi \|_\HN{s+2}
\end{align*}
On the other hand in Section \ref{sec:convergence} we had:
\begin{align*}
&\|W_\varepsilon^l (\phi)-T(\varepsilon l)(\phi) \|_{\HN{s}}\leq \varepsilon^2 l C ||\ket{\phi}||_\HN{s+2}
\end{align*}
And thus the bound on the overall error is: 
\begin{align*}
&\|\RC(W_\varepsilon^l (\D(\phi)))-T(\varepsilon l)\phi \|_{\HN{s}}\\
&\leq \varepsilon^2 (lC+C') ||\ket{\phi}||_\HN{s+2}.
\end{align*}
where in the last inequality we should recall that $\varepsilon$ is the discretization parameter and $k$ the number of iterations, thus $x_0=\varepsilon l$ is for how long the evolution is simulated.

\section{Further considerations}\label{sec:considerations}

\noindent {\em Generalizations.} The method would work equally well for any symmetric hyperbolic systems with rational eigenvalues, i.e. equations of the form
\begin{align*}
\ii\partial_0 \ket{\psi} &= D\ket{\psi}\quad \textrm{with}\quad D=\beta^0-\ii\sum_j \beta^j\partial_j
\end{align*}
where the $(\beta^j)$ are $d\times d$ hermitian having rational eigenvalues, and $\beta^0$ is hermitian.  We can write $\beta^j=\frac{1}{q}(U_j)\Delta^j(U_j)^{\dagger}$, with $q\in\N^*$, $U_j$ unitary, and $\Delta^j$ diagonal with integer coefficients $\lambda_0^j,\ldots,\lambda_d^j$.  The same procedure yields the quantum walk:
\begin{align*}
&W_\varepsilon=C_\varepsilon \prod_j U_jT_{j,\varepsilon}U_j^\dagger\quad \textrm{with}\\
&C_\varepsilon=\exp{(-\ii\varepsilon q\beta^0)}\quad\textrm{and}\quad T_{j,\varepsilon}=\sum_{r,l} |r\rangle\bra{l}\tr{j,-\lambda^j_r\varepsilon}
\end{align*}
More generally even, the method would work for equations of the form
\begin{align*}
\ii\partial_0 \ket{\psi} &= D\ket{\psi}\quad \textrm{with}\\
D&=\sum_j D_j
\end{align*}
such that each $\exp{(-\ii D_j)}$ is a quantum walk. Indeed, the same procedure yields the quantum walk
$$W=\prod_j \exp\pa{-\ii D_j}.$$
Ultimately, it is the fact the Dirac Hamiltonian is a sum of logarithms of Quantum Walks, which enables us to model it as the product of these Quantum Walks.

\noindent {\em Observational equivalence.} Consider the case when $s=0$. We then have $\LS=\HS{0}$, as $||.||_2=||.||_\HN{0}$: the Sobolev norm then coincides with that of quantum theory, and we can interpret convergence in an operational manner. Convergence gives us the existence of $C$ such that if $\ket{\psi(x_0=0)}\in \HS{2}$, then for all $\varepsilon_l=x_0/l$ we have
$$||W_{\varepsilon_l}^l\ket{\psi(0)} - \ket{\psi(x_0)}||_2 \leq \varepsilon x_0 C||\ket{\psi(0)}||_\HN{2}.$$
According to quantum theory the probability of observing through a measurement a discrepancy between the iterated walk $W_{\varepsilon_l}^l\ket{\phi}$ and the solution of the Dirac equation $\ket{\psi(x_0)}$ is given by $\sin^2(\theta)$, with $\theta$ the angle between both vectors. Simple trigonometric reasoning shows that this is bounded above by $\varepsilon^2 x_0^2 C^2||\ket{\psi(0)}||^2$, i.e. it diminishes quadratically as $\varepsilon$ goes to zero.

\section*{Summary}

\noindent The Quantum Walk
\begin{align*}
W_\varepsilon=C_\varepsilon(\Id\otimes H)T_{1,\varepsilon}(\Id \otimes HF)T_{2,\varepsilon}(\Id\otimes F^\dagger) T_{3,\varepsilon}
\end{align*}
models the Dirac equation. Indeed, consistency is ensured to first order and stability is given by unitarity, hence the model is convergent to first order. The result can be specialized elegantly to lower dimensions. It can also be generalized to other first-order PDEs, as well as to PDEs whose Hamiltonians can be expressed as a sum of logarithms of Quantum Walks. The model is suitable for quantum simulation, or as a discrete toy model.
\noindent The Quantum Walk is parametrized on $\varepsilon$, the discretization step. It is of course tempting to set $\varepsilon$ to in Planck units, and grant
\begin{align*}
W=C(\Id\otimes H)T_{1}(\Id\otimes HF)T_{2}(\Id\otimes F^\dagger) T_{3}
\end{align*}
a more fundamental status. One could even wonder whether some relativistic particles might behave according to this Quantum Walk, rather than the Dirac equation. To our reader, we ask: could experimentalists really tell the difference?\\
A decohered version of the quantum walk model could be studied using the general techniques of \cite{dissipative_CT}.
We plan to study to which extent such discrete models retain some Poincar\'e-invariance.

\section*{Acknowledgements} The authors are indebted to Olivier Bournez and David Meyer for insightful discussions at the early stages of this work, to St\'{e}phane Labb\'{e} and Stefano Facchini for some advices, and Alain Joye for his help.

\bibliography{biblio}

\appendix

\section{Facts in Fourier space}\label{sec:Fourier}
	
\noindent {\em Fourier transform.} We recall that the Fourier transform of the wave-function $\phi \in \LS$ is defined as the function $(\FT \ket{\phi})= \hat{\phi} : \R^n \rightarrow \C^d$ such that
\begin{equation*}
\ket{\hat{\phi}} ( k) = \frac{1}{(2\pi)^{n/2}} \int_{\R^n} \ket{\phi(x)}e^{-i k \cdot x}~\dd x
\end{equation*}
where by $ k \cdot x$ we mean the scalar product in Euclidean space $\R^n$, $x=(x_j)$, and $ k=( k_j)$. The function $\FT$ is unitary, its inverse is
\begin{equation*}
\ket{\phi} (x) = \frac{1}{(2\pi)^{n/2}} \int_{\R^n} \ket{\hat{\phi}( k)}e^{i k \cdot x}~\dd  k.
\end{equation*}
From the above definition it is easily seen that for the spatial derivatives: $\FT (\partial_j \phi) ( k) = i  k_j \hat{\phi}$. Is is also useful to recall that for translations: 
\begin{align*}
\FT \left( \phi (x \pm \varepsilon) \right) ( k) &= e^{\pm \ii  k \cdot \varepsilon}~\hat{\phi}( k)
\end{align*}	

In Fourier space the $(2+1)$-dimensional Dirac operator, Eq. (\ref{eq:Dirac2D}), becomes:
\begin{align*}
\hat{D}( k) 
&= m\sigma^2 +  k_1 \sigma^1 +  k_2 \sigma^3 
\\ &=\left(\begin{array}{cc}
 k_2 & k_1-\ii m\\
 k_1+ \ii m  &- k_2
\end{array}\right)
\end{align*}
with eigenvalues $\pm |\gamma|$, being $\gamma^2 = m^2+|| k||^2$. The same formula for the eigenvalues holds true in three dimensions (i.e. there is a twofold degeneracy).

In Fourier space the $(2+1)$-dimensional Dirac Quantum Walk operator $\hat{W}_\varepsilon$, decomposes as a product of exponential matrices, using identities such as: 
\begin{align*}
H \hat{T}_{1,\varepsilon}( k) H &=  H \left(\begin{array}{cc}
e^{-\ii  k_1 \varepsilon} & 0 \\
0 & e^{ \ii  k_1 \varepsilon} 
\end{array}\right) H \\ &= He^{-\ii  k_1  \varepsilon \sigma^3}H = e^{-\ii \varepsilon  k_1 \sigma^1}
\end{align*}
and likewise for the other directions. Eventually in $(n+1)$-dimensions it takes the form
$$
\hat{W}_\varepsilon = \prod_{\mu} e^{-\ii \varepsilon \hat{A}_\mu }
$$
with some known $\hat{A}_\mu$. 

\noindent {\em Fourier series.} We recall that the Fourier series of the wave-function $\phi \in L^2([-\frac{\pi}{\varepsilon},\frac{\pi}{\varepsilon}]^n)^d$, $\varepsilon \in \R^{+}$, is defined as the function $(\FS \ket{\phi})= \hat{\phi} : \varepsilon\Z^n \rightarrow \C^d$ such that
\begin{equation*}
\ket{\hat{\phi}} ( k) = \left(\frac{\varepsilon}{2\pi}\right)^{n/2} \int_{[-\frac{\pi}{\varepsilon},\frac{\pi}{\varepsilon}]^n} \ket{\phi(x)}e^{i k \cdot x}~\dd x.
\end{equation*}
The function $\FS$ is unitary, its inverse is
\begin{equation*}
\ket{\phi} (x) = \left( \frac{\varepsilon}{2\pi}\right)^{n/2} \sum_{ k\in \varepsilon\Z^n} \ket{\hat{\phi}( k)}e^{-i k \cdot x}.
\end{equation*}
The sign conventions of the exponentials are non-standard; they have been chosen to that, whenever $\hat{\phi}=\FT(\phi)$ has support in $[-\frac{\pi}{\varepsilon},\frac{\pi}{\varepsilon}]^n$, then (with $|_X$ denoting restriction to $X$):
\begin{itemize}
\item $\FS(\hat\phi|_{[-\frac{\pi}{\varepsilon},\frac{\pi}{\varepsilon}]^n}) =\varepsilon^{n/2}\FT^{-1}(\hat\phi)|_{\varepsilon\Z^n}=\varepsilon^{n/2}\phi|_{\varepsilon\Z^n}$;
\item $\FS^{-1}(\varepsilon^{n/2}\phi|_{\varepsilon\Z^n}) =\hat\phi|_{[-\frac{\pi}{\varepsilon},\frac{\pi}{\varepsilon}]^n}$.
\end{itemize}
Indeed, the first point follows from the definition, and the second is the reciprocal.

\section{Sobolev spaces and Well-posedness}\label{sec:Sobolev}

{\em Sobolev spaces.} The usual wave-function space for quantum theory is the subspace $\LS$ of the functions $\mathbb{R}^n\rightarrow \mathbb{C}^d$ for which the $||.||_2$-norm is finite. Recall that
\begin{align*}
||\ket{\phi}||_2&=\sqrt{ \int_{\mathbb{R}^n} ||\ket{\phi(x)}||^2 \dd x}
\end{align*}
with $||.||$ the usual $2$-norm in $\mathbb{C}^d$, $x=(x_j)$. For our approximations to hold, we need to restrict to the subspace $\HS{s}$ of the functions $\LS$ for which the $||.||_\HN{s}$-norm is finite.  Recall that
\begin{align*}
||\ket{\phi}||_\HN{s}&=\sqrt{ \int_{\mathbb{R}^n} (1+m^2+|| k||^2)^{s}||\ket{\hat{\phi}( k)}||^2 \dd k}
\end{align*}
with $\ket{\hat{\phi}}$ the Fourier transform of $\ket{\phi}$, and again $|| k||^2=\sum_j | k_j|^2$. \\
Several remarks are in order. First, notice that
$||\ket{\phi}||_\HN{0}=||\ket{\hat{\phi}}||_2=||\ket{\phi}||_2$, thus $\HS{0}=\LS$. Second, notice that for continuous differentiable functions, $||\ket{\phi}||^2_\HN{1}=(1+m^2)||\ket{\phi}||^2_2+\sum_j||\partial_j\ket{\phi}||^2_2$, thus $\HS{1}$ is just the subset of $\LS$ having first-order derivatives in $\LS$. The same holds for $\HS{s+1}$ with respect to $\HS{s}$. Third, notice that $\HS{s+1}$ is dense in $\HS{s}$, as can be seen from mollification techniques \cite{sobolev_original}. Finally, notice that, on the one hand, the choice of having the $||.||_\HN{s}$-norm to depend on $m$ is slightly non-standard: usually this constant is set to zero.\\
On the other hand, three elements argue in favour of this non-standard choice: 1/ this fits nicely with the mathematics of this paper; 2/ our main use of the $||.||_\HN{s>0}$-norm is to impose a sufficiently regular initial condition on the particle's wave-function, that this regularity condition may depend on the particle's mass $m$ does not seem problematic; 3/ the above defined $||.||_\HN{s}$-norm is equivalent to the usual $||.||_{H^s}$-norm: 
$$||\ket{\phi}||_{H^s} = \sqrt{\int_{\mathbb{R}^n} (1+|| k||^2)^{s}||\ket{\hat{\phi}( k)}||^2 d k}$$ in the sense of norm equivalence, because $1+|| k||^2\leq 1+m^2+|| k||^2 \leq (m^2+1)(1+|| k||^2)$ . This last point is why the well-posedness of the Dirac equation with respect to the usual $||.||_{H^s}$-norm carries through with respect to the $||.||_\HN{s}$-norm, see next.

\noindent {\em Well-posedness.} A Cauchy problem $\partial_0 \ket{\psi}=D\ket{\psi}$ is well-posed in a Banach space $X$ if:
\begin{itemize}
\item $D$ is a densely defined operator of $X$;

\item There exists a dense subset $Y$ of $X$ such that for every initial condition in $Y$, the Cauchy problem has a solution;

\item There exists a non-decreasing function $C:\R^+\to\R^+$ such that for every solution $\ket{\psi}$ (not necessarily from an initial condition in $Y$) and every $x_0\in \R^+$, $||\ket{\psi(x_0)}||_X\leq C(x_0) ||\ket{\psi(0)}||_X$. 

\end{itemize}

A hyperbolic symmetric system is a Cauchy problem of the 
form
$$ \partial_0 \ket{\psi} = D\ket{\psi}\quad \textrm{with}\quad D=-\ii\beta^0-\sum_j \beta^j\partial_j $$
where the $(\beta^\mu)$ are hermitian.

For symmetric hyperbolic systems, the Cauchy problem is known to be well-posed in $H^s({\mathbb{R}}^n)^d$ for any $s\geq 0$.  $D$ is defined on the subspace of $\ket{\phi}\in H^s({\mathbb{R}}^n)^d$ such that $D\ket{\psi}\in H^s({\mathbb{R}}^n)^d$, which is dense indeed, and every initial condition in this space yields a solution.  The $H^s$-norm is constant for solutions of the problem, so that $C(t)=1$ fulfills the requirement.  For references, see \cite{Fattorini} (1.6.21) or \cite{Benzoni}.

\couic{

where:
\begin{itemize}
\item the notation $\ket{\psi^\varepsilon}= O(\varepsilon^2)$ refers to a property of the function $\varepsilon\mapsto\ket{\psi^\varepsilon}$. It states that 
states that there exists $M>0$ and $\delta>0$ such that for all $|\varepsilon|<\delta$, we have $||\ket{\psi^\varepsilon}||\leq M\varepsilon^2$. This is w.r.t. the $L_2$-norm.
\item the notation $E^\varepsilon= O(\varepsilon^2)$ refers to a property of the function $\varepsilon\mapsto E^\varepsilon$. It states that there exists $M>0$ and $\delta>0$ such that for all $|\varepsilon|<\delta$ and $\ket{\psi}$, we have that $||E^\varepsilon \ket{\psi}||\leq M||\ket{\psi}||\varepsilon^2$. I.e. the same property in operator norm.
\end{itemize}

}

\end{document}

%% file: kkmacros.tex
\usepackage{amsmath,amsfonts,amssymb,bbm}

\newcommand{\bra}[1]{\langle #1 |}

\newcommand{\pa}[1]{\left(#1\right)}
\newcommand{\ket}[1]{#1} 

\newcommand{\couic}[1]{}

\newcommand{\D}{\textrm{Discretize}}
\newcommand{\RN}{\textrm{Renormalize}}
\newcommand{\RC}{\textrm{Reconstruct}}
\newcommand{\FT}{\mathcal{FT}}
\newcommand{\FS}{\mathcal{FS}}

\newcommand{\dd}{\mathrm{d}}

\newcommand{\C}{\mathbb{C}}
\newcommand{\N}{\mathbb{N}}
\newcommand{\R}{\mathbb{R}}

\newcommand{\Z}{\mathbb{Z}}

\newcommand{\Id}{\operatorname{Id}}

\newcommand{\LS}{L^2(\mathbb{R}^n)^d}
\newcommand{\HS}[1]{H^{#1}_m(\mathbb{R}^n)^d}
\newcommand{\HN}[1]{{H^{#1}_m}}

\newcommand{\tr}[1]{\tau_{#1}}
\newcommand{\ii}{\mathrm i}

%% file: dirac5.bbl
\begin{thebibliography}{38}
\expandafter\ifx\csname natexlab\endcsname\relax\def\natexlab#1{#1}\fi
\expandafter\ifx\csname bibnamefont\endcsname\relax
  \def\bibnamefont#1{#1}\fi
\expandafter\ifx\csname bibfnamefont\endcsname\relax
  \def\bibfnamefont#1{#1}\fi
\expandafter\ifx\csname citenamefont\endcsname\relax
  \def\citenamefont#1{#1}\fi
\expandafter\ifx\csname url\endcsname\relax
  \def\url#1{\texttt{#1}}\fi
\expandafter\ifx\csname urlprefix\endcsname\relax\def\urlprefix{URL }\fi
\providecommand{\bibinfo}[2]{#2}
\providecommand{\eprint}[2][]{\url{#2}}

\bibitem[{\citenamefont{Feynman}(1982)}]{FeynmanQC}
\bibinfo{author}{\bibfnamefont{R.~P.} \bibnamefont{Feynman}},
  \bibinfo{journal}{International Journal of Theoretical Physics}
  \textbf{\bibinfo{volume}{21}}, \bibinfo{pages}{467} (\bibinfo{year}{1982}).

\bibitem[{\citenamefont{Succi and Benzi}(1993)}]{BenziSucci}
\bibinfo{author}{\bibfnamefont{S.}~\bibnamefont{Succi}} \bibnamefont{and}
  \bibinfo{author}{\bibfnamefont{R.}~\bibnamefont{Benzi}},
  \bibinfo{journal}{Physica D: Nonlinear Phenomena}
  \textbf{\bibinfo{volume}{69}}, \bibinfo{pages}{327} (\bibinfo{year}{1993}).

\bibitem[{\citenamefont{Bialynicki-Birula}(1994)}]{Bialynicki-Birula}
\bibinfo{author}{\bibfnamefont{I.}~\bibnamefont{Bialynicki-Birula}},
  \bibinfo{journal}{Phys. Rev. D.} \textbf{\bibinfo{volume}{49}},
  \bibinfo{pages}{6920} (\bibinfo{year}{1994}).

\bibitem[{\citenamefont{Meyer}(1996)}]{MeyerQLGI}
\bibinfo{author}{\bibfnamefont{D.~A.} \bibnamefont{Meyer}},
  \bibinfo{journal}{J. Stat. Phys} \textbf{\bibinfo{volume}{85}},
  \bibinfo{pages}{551} (\bibinfo{year}{1996}).

\bibitem[{\citenamefont{Kempe}(2003)}]{Kempe}
\bibinfo{author}{\bibfnamefont{J.}~\bibnamefont{Kempe}},
  \bibinfo{journal}{Contemporary Physics} \textbf{\bibinfo{volume}{44}},
  \bibinfo{pages}{307} (\bibinfo{year}{2003}).

\bibitem[{\citenamefont{Chandrashekar et~al.}(2010)\citenamefont{Chandrashekar,
  Banerjee, and Srikanth}}]{IndiansDirac}
\bibinfo{author}{\bibfnamefont{C.}~\bibnamefont{Chandrashekar}},
  \bibinfo{author}{\bibfnamefont{S.}~\bibnamefont{Banerjee}}, \bibnamefont{and}
  \bibinfo{author}{\bibfnamefont{R.}~\bibnamefont{Srikanth}},
  \bibinfo{journal}{Phys. Rev. A.} \textbf{\bibinfo{volume}{81}},
  \bibinfo{pages}{62340} (\bibinfo{year}{2010}), ISSN
  \bibinfo{issn}{1094-1622}.

\bibitem[{\citenamefont{Love and Boghosian}(2005)}]{LoveBoghosian}
\bibinfo{author}{\bibfnamefont{P.}~\bibnamefont{Love}} \bibnamefont{and}
  \bibinfo{author}{\bibfnamefont{B.}~\bibnamefont{Boghosian}},
  \bibinfo{journal}{Quantum Information Processing}
  \textbf{\bibinfo{volume}{4}}, \bibinfo{pages}{335} (\bibinfo{year}{2005}),
  ISSN \bibinfo{issn}{1570-0755}.

\bibitem[{\citenamefont{Palpacelli}(2009)}]{PalpacelliThesis}
\bibinfo{author}{\bibfnamefont{S.}~\bibnamefont{Palpacelli}},
  \bibinfo{type}{Ph.D. thesis}, \bibinfo{institution}{Università Degli Studi
  Roma Tre, Facoltà di Scienze Matematiche Fisiche e Naturali, Dottorato in
  Matematica XXI ciclo} (\bibinfo{year}{2009}).

\bibitem[{\citenamefont{Lapitski and Dellar}(2011)}]{LapitskiDellar}
\bibinfo{author}{\bibfnamefont{D.}~\bibnamefont{Lapitski}} \bibnamefont{and}
  \bibinfo{author}{\bibfnamefont{P.~J.} \bibnamefont{Dellar}},
  \bibinfo{journal}{Philosophical Transactions of the Royal Society A:
  Mathematical, Physical and Engineering Sciences}
  \textbf{\bibinfo{volume}{369}}, \bibinfo{pages}{2155} (\bibinfo{year}{2011}).

\bibitem[{\citenamefont{Dellar et~al.}(2011)\citenamefont{Dellar, Lapitski,
  Palpacelli, and Succi}}]{LapitskiDellarPalpacelliSucci}
\bibinfo{author}{\bibfnamefont{P.~J.} \bibnamefont{Dellar}},
  \bibinfo{author}{\bibfnamefont{D.}~\bibnamefont{Lapitski}},
  \bibinfo{author}{\bibfnamefont{S.}~\bibnamefont{Palpacelli}},
  \bibnamefont{and} \bibinfo{author}{\bibfnamefont{S.}~\bibnamefont{Succi}},
  \bibinfo{journal}{Phys. Rev. E} \textbf{\bibinfo{volume}{83}},
  \bibinfo{pages}{046706} (\bibinfo{year}{2011}),
  \urlprefix\url{http://link.aps.org/doi/10.1103/PhysRevE.83.046706}.

\bibitem[{\citenamefont{Strauch}(2006{\natexlab{a}})}]{StrauchShrodinger}
\bibinfo{author}{\bibfnamefont{F.~W.} \bibnamefont{Strauch}},
  \bibinfo{journal}{Physical Review. A} \textbf{\bibinfo{volume}{73}}
  (\bibinfo{year}{2006}{\natexlab{a}}).

\bibitem[{\citenamefont{Strauch}(2007)}]{StrauchPhenomena}
\bibinfo{author}{\bibfnamefont{F.}~\bibnamefont{Strauch}},
  \bibinfo{journal}{Journal of Mathematical Physics}
  \textbf{\bibinfo{volume}{48}}, \bibinfo{pages}{082102}
  (\bibinfo{year}{2007}).

\bibitem[{\citenamefont{Bisio et~al.}(2012)\citenamefont{Bisio, D'Ariano, and
  Tosini}}]{DAriano}
\bibinfo{author}{\bibfnamefont{A.}~\bibnamefont{Bisio}},
  \bibinfo{author}{\bibfnamefont{G.~M.} \bibnamefont{D'Ariano}},
  \bibnamefont{and} \bibinfo{author}{\bibfnamefont{A.}~\bibnamefont{Tosini}},
  \bibinfo{journal}{arXiv preprint arXiv:1212.2839}  (\bibinfo{year}{2012}).

\bibitem[{\citenamefont{Strauch}(2006{\natexlab{b}})}]{StrauchCTQW}
\bibinfo{author}{\bibfnamefont{F.~W.} \bibnamefont{Strauch}},
  \bibinfo{journal}{Physical Review A} \textbf{\bibinfo{volume}{74}},
  \bibinfo{pages}{030301} (\bibinfo{year}{2006}{\natexlab{b}}).

\bibitem[{\citenamefont{Boghosian and
  Taylor}(1998{\natexlab{a}})}]{BoghosianTaylor1}
\bibinfo{author}{\bibfnamefont{B.~M.} \bibnamefont{Boghosian}}
  \bibnamefont{and} \bibinfo{author}{\bibfnamefont{W.}~\bibnamefont{Taylor}},
  \bibinfo{journal}{Physica D} \textbf{\bibinfo{volume}{120}},
  \bibinfo{pages}{30} (\bibinfo{year}{1998}{\natexlab{a}}).

\bibitem[{\citenamefont{Cha}(2011)}]{ChaMeyer}
\bibinfo{author}{\bibfnamefont{M.}~\bibnamefont{Cha}}, Master's thesis,
  \bibinfo{school}{University of California}, \bibinfo{address}{San Diego}
  (\bibinfo{year}{2011}).

\bibitem[{\citenamefont{Boghosian and
  Taylor}(1998{\natexlab{b}})}]{BoghosianTaylor2}
\bibinfo{author}{\bibfnamefont{B.~M.} \bibnamefont{Boghosian}}
  \bibnamefont{and} \bibinfo{author}{\bibfnamefont{W.}~\bibnamefont{Taylor}},
  \bibinfo{journal}{Phys. Rev. E.} \textbf{\bibinfo{volume}{57}},
  \bibinfo{pages}{54} (\bibinfo{year}{1998}{\natexlab{b}}).

\bibitem[{\citenamefont{Lorin and Bandrauk}(2011)}]{LorinBandrauk}
\bibinfo{author}{\bibfnamefont{E.}~\bibnamefont{Lorin}} \bibnamefont{and}
  \bibinfo{author}{\bibfnamefont{A.}~\bibnamefont{Bandrauk}},
  \bibinfo{journal}{Nonlinear Analysis: Real World Applications}
  \textbf{\bibinfo{volume}{12}}, \bibinfo{pages}{190} (\bibinfo{year}{2011}).

\bibitem[{\citenamefont{Huang et~al.}(2005)\citenamefont{Huang, Jin, Markowich,
  Sparber, and Zheng}}]{HJMSZ}
\bibinfo{author}{\bibfnamefont{Z.}~\bibnamefont{Huang}},
  \bibinfo{author}{\bibfnamefont{S.}~\bibnamefont{Jin}},
  \bibinfo{author}{\bibfnamefont{P.~A.} \bibnamefont{Markowich}},
  \bibinfo{author}{\bibfnamefont{C.}~\bibnamefont{Sparber}}, \bibnamefont{and}
  \bibinfo{author}{\bibfnamefont{C.}~\bibnamefont{Zheng}},
  \bibinfo{journal}{Journal of Computational Physics}
  \textbf{\bibinfo{volume}{208}}, \bibinfo{pages}{761} (\bibinfo{year}{2005}).

\bibitem[{\citenamefont{Fillion-Gourdeau
  et~al.}(2012)\citenamefont{Fillion-Gourdeau, Lorin, and
  Bandrauk}}]{FillionLorinBandrauk}
\bibinfo{author}{\bibfnamefont{F.}~\bibnamefont{Fillion-Gourdeau}},
  \bibinfo{author}{\bibfnamefont{E.}~\bibnamefont{Lorin}}, \bibnamefont{and}
  \bibinfo{author}{\bibfnamefont{A.~D.} \bibnamefont{Bandrauk}},
  \bibinfo{journal}{Computer Physics Communications}
  \textbf{\bibinfo{volume}{183}}, \bibinfo{pages}{1403} (\bibinfo{year}{2012}).

\bibitem[{\citenamefont{Childs and Goldstone}(2004)}]{ChildsGoldstone}
\bibinfo{author}{\bibfnamefont{A.~M.} \bibnamefont{Childs}} \bibnamefont{and}
  \bibinfo{author}{\bibfnamefont{J.}~\bibnamefont{Goldstone}},
  \bibinfo{journal}{Physical Review A} \textbf{\bibinfo{volume}{70}},
  \bibinfo{pages}{042312} (\bibinfo{year}{2004}).

\bibitem[{\citenamefont{D'Ariano and Perinotti}(2013)}]{DAriano3D}
\bibinfo{author}{\bibfnamefont{G.~M.} \bibnamefont{D'Ariano}} \bibnamefont{and}
  \bibinfo{author}{\bibfnamefont{P.}~\bibnamefont{Perinotti}}
  (\bibinfo{year}{2013}), \bibinfo{note}{pre-print arXiv:1306.1934}.

\bibitem[{\citenamefont{Bateson}(2012)}]{Bateson2012}
\bibinfo{author}{\bibfnamefont{R.}~\bibnamefont{Bateson}}, in
  \emph{\bibinfo{booktitle}{Journal of Physics: Conference Series}}
  (\bibinfo{organization}{IOP Publishing}, \bibinfo{year}{2012}), vol.
  \bibinfo{volume}{361}, p. \bibinfo{pages}{012009}.

\bibitem[{\citenamefont{Kauffman and Noyes}(1996)}]{Kauffman}
\bibinfo{author}{\bibfnamefont{L.~H.} \bibnamefont{Kauffman}} \bibnamefont{and}
  \bibinfo{author}{\bibfnamefont{H.~P.} \bibnamefont{Noyes}},
  \bibinfo{journal}{Physics Letters A} \textbf{\bibinfo{volume}{218}},
  \bibinfo{pages}{139} (\bibinfo{year}{1996}), ISSN \bibinfo{issn}{0375-9601}.

\bibitem[{\citenamefont{Gersch}(1981)}]{Gersch_chessboard}
\bibinfo{author}{\bibnamefont{Gersch}}, \bibinfo{journal}{Int. J. Theo. Phys.}
  \textbf{\bibinfo{volume}{20}}, \bibinfo{pages}{491} (\bibinfo{year}{1981}),
  \bibinfo{note}{feynman relativistic chessboard}.

\bibitem[{\citenamefont{Kishigi et~al.}(2008)\citenamefont{Kishigi, Takeda, and
  Hasegawa}}]{japanese_honeycomb}
\bibinfo{author}{\bibfnamefont{K.}~\bibnamefont{Kishigi}},
  \bibinfo{author}{\bibfnamefont{R.}~\bibnamefont{Takeda}}, \bibnamefont{and}
  \bibinfo{author}{\bibfnamefont{Y.}~\bibnamefont{Hasegawa}},
  \bibinfo{journal}{Journal of Physics: Conference Series}
  \textbf{\bibinfo{volume}{132}}, \bibinfo{pages}{012005}
  (\bibinfo{year}{2008}).

\bibitem[{\citenamefont{Di~Molfetta and Debbasch}(2011)}]{MolfettaDebbasch}
\bibinfo{author}{\bibfnamefont{G.}~\bibnamefont{Di~Molfetta}} \bibnamefont{and}
  \bibinfo{author}{\bibfnamefont{F.}~\bibnamefont{Debbasch}},
  \bibinfo{journal}{arXiv preprint arXiv:1111.2165}  (\bibinfo{year}{2011}).

\bibitem[{\citenamefont{Di~Franco et~al.}(2011)\citenamefont{Di~Franco,
  Mc~Gettrick, Machida, and Busch}}]{McGettrick}
\bibinfo{author}{\bibfnamefont{C.}~\bibnamefont{Di~Franco}},
  \bibinfo{author}{\bibfnamefont{M.}~\bibnamefont{Mc~Gettrick}},
  \bibinfo{author}{\bibfnamefont{T.}~\bibnamefont{Machida}}, \bibnamefont{and}
  \bibinfo{author}{\bibfnamefont{T.}~\bibnamefont{Busch}},
  \bibinfo{journal}{Physical Review A} \textbf{\bibinfo{volume}{84}},
  \bibinfo{pages}{042337} (\bibinfo{year}{2011}).

\bibitem[{\citenamefont{Fattorini}(1983)}]{Fattorini}
\bibinfo{author}{\bibfnamefont{H.~O.} \bibnamefont{Fattorini}},
  \emph{\bibinfo{title}{The {C}auchy Problem}}, no.~\bibinfo{number}{18} in
  \bibinfo{series}{Encyclopedia of Mathematics and its Applications}
  (\bibinfo{publisher}{Cambridge}, \bibinfo{year}{1983}).

\bibitem[{Tay(2003)}]{TaylorIF}
 (\bibinfo{year}{2003}),
  \urlprefix\url{http://www.math.binghamton.edu/loya/papers/kl_taylor.pdf}.

\bibitem[{\citenamefont{Davies}(2007)}]{DaviesLinear}
\bibinfo{author}{\bibfnamefont{E.~B.} \bibnamefont{Davies}},
  \emph{\bibinfo{title}{Linear operators and their spectra}}, vol.
  \bibinfo{volume}{106} (\bibinfo{publisher}{Cambridge University Press},
  \bibinfo{year}{2007}).

\bibitem[{\citenamefont{Lax and Richtmyer}(1956)}]{LaxRichtmyer}
\bibinfo{author}{\bibfnamefont{P.~D.} \bibnamefont{Lax}} \bibnamefont{and}
  \bibinfo{author}{\bibfnamefont{R.~D.} \bibnamefont{Richtmyer}},
  \bibinfo{journal}{Communications on Pure and Applied Mathematics}
  \textbf{\bibinfo{volume}{9}}, \bibinfo{pages}{267} (\bibinfo{year}{1956}).

\bibitem[{ISE(2011--2012)}]{ISEM-Lax}
 (\bibinfo{year}{2011--2012}), \bibinfo{note}{15th Internet Seminar, Operator
  Semigroups for Numerical Analysis},
  \urlprefix\url{https://isem-mathematik.uibk.ac.at/isemwiki/index.php/Lecture%
_4}.

\bibitem[{\citenamefont{Petersen and Middleton}(1962)}]{pm62}
\bibinfo{author}{\bibfnamefont{D.~P.} \bibnamefont{Petersen}} \bibnamefont{and}
  \bibinfo{author}{\bibfnamefont{D.}~\bibnamefont{Middleton}},
  \bibinfo{journal}{Information and Control} \textbf{\bibinfo{volume}{5}},
  \bibinfo{pages}{279} (\bibinfo{year}{1962}).

\bibitem[{\citenamefont{Jingfan and Gensun}(2004)}]{sampling_chinois}
\bibinfo{author}{\bibfnamefont{L.}~\bibnamefont{Jingfan}} \bibnamefont{and}
  \bibinfo{author}{\bibfnamefont{F.}~\bibnamefont{Gensun}},
  \bibinfo{journal}{Analysis in Theory and Applications}
  \textbf{\bibinfo{volume}{20}}, \bibinfo{pages}{52} (\bibinfo{year}{2004}),
  ISSN \bibinfo{issn}{1672-4070},
  \urlprefix\url{http://dx.doi.org/10.1007/BF02835258}.

\bibitem[{\citenamefont{Kliesch et~al.}(2011)\citenamefont{Kliesch, Barthel,
  Gogolin, Kastoryano, and Eisert}}]{dissipative_CT}
\bibinfo{author}{\bibfnamefont{M.}~\bibnamefont{Kliesch}},
  \bibinfo{author}{\bibfnamefont{T.}~\bibnamefont{Barthel}},
  \bibinfo{author}{\bibfnamefont{C.}~\bibnamefont{Gogolin}},
  \bibinfo{author}{\bibfnamefont{M.}~\bibnamefont{Kastoryano}},
  \bibnamefont{and} \bibinfo{author}{\bibfnamefont{J.}~\bibnamefont{Eisert}},
  \bibinfo{journal}{Phys. Rev. Lett.} \textbf{\bibinfo{volume}{107}},
  \bibinfo{pages}{120501} (\bibinfo{year}{2011}).

\bibitem[{\citenamefont{Sobolev}(1938)}]{sobolev_original}
\bibinfo{author}{\bibfnamefont{S.}~\bibnamefont{Sobolev}},
  \bibinfo{journal}{Rec. Math. [Mat. Sbornik] N.S.}
  \textbf{\bibinfo{volume}{4}}, \bibinfo{pages}{471} (\bibinfo{year}{1938}).

\bibitem[{\citenamefont{Benzoni-Gavage and Serre}(2007)}]{Benzoni}
\bibinfo{author}{\bibfnamefont{S.}~\bibnamefont{Benzoni-Gavage}}
  \bibnamefont{and} \bibinfo{author}{\bibfnamefont{D.}~\bibnamefont{Serre}},
  \emph{\bibinfo{title}{Multi-dimensional hyperbolic partial differential
  equations}} (\bibinfo{publisher}{Oxford University Press},
  \bibinfo{year}{2007}).

\end{thebibliography}
